# Organisational justice moderates the link between leadership, work engagement and innovation work behaviour

**Authors:**
Rahmat Sabuhari[1]
Rusman Soleman[2]
Marwan Man Soleman[1]
Johan Fahri[1]
Muhammad Rachmat[1]

**Affiliations:**
[1]Department of Management, Faculty of Economics and Business, Universitas Khairun, Ternate, Indonesia

[2]Department of Accountancy, Faculty of Economics and Business, Universitas Khairun, Ternate, Indonesia

**Corresponding author:**
Rahmat Sabuhari,
rahmat.sabuhari@unkhair.ac.id

**Background:** Regional civil servants in Indonesia, especially in the eastern regions, are now required to exhibit innovative work behaviour (IWB) as mandated by national regulations. While transformational leadership (TFL) and work engagement (WE) are believed to foster such behaviour, the impact and the moderating role of organisational justice (OJ) remain underexplored. This study addressed that gap by examining how TFL and WE contribute to IWB development within the context of public sector work, with OJ potentially moderating these relationships.

**Aim:** This study aimed to determine the direct influence of TFL, WE and OJ on IWB and to analyse the role of OJ as a moderating variable.

**Setting:** This research was conducted in the eastern region of Indonesia, specifically North Maluku province, in various regional apparatus organisations, and a random sample of 500 people was taken.

**Method:** This study used structural equation modelling-partial least squares (SEM-PLS) regression analysis as a data analysis method to test the causal relationship and moderation effects of each hypothesis.

**Results:** The study findings reveal that while TFL and WE directly influence IWB, OJ neither has a direct effect nor serves as a moderating variable.

**Conclusion:** These findings offer theoretical and practical contributions for Indonesian local governments, emphasising the importance of leadership and engagement in enhancing innovative behaviour at work.

**Contribution:** This study advances social exchange theory through civil servants' innovative behaviour, exploring how TFL, WE, and OJ drive innovation. It also empirically tests OJ as a moderator in the relationship between TFL, WE, and IWB.

## Introduction

Rising citizen expectations, limited resources and changing social, environmental and economic conditions are driving governments around the world to continue to find new and innovative ways to provide public services (Pooe & Munyanyi 2022; Taylor 2018). In Indonesia, the government encourages regional creativity and innovation by measuring and assessing regional innovation indexes and awarding regional government innovation awards issued by the Ministry of Home Affairs of the Republic of Indonesia in 2022 and 2024. In 2022 North Maluku province had an index value of 33.11, which increased to an index value of 58.47 in 2024. This shows that innovative work behaviour (IWB) in the local government sector has not been in line with the expectations of the central government. The government's hope for increasing regional innovation is to accelerate the achievement of community welfare, increase regional competitiveness and strengthen regional independence by improving public services, community empowerment and quality of life. Regions that are declared to have a very innovative index if they have a score of 60–100. Innovation is crucial for Regional Work Units (RWUs) to remain competitive and effective, specifically referring to the definition outlined in Government Regulation Number 38 of 2017 on Regional Innovation, which encompasses all forms of change in local government implementation aimed at enhancing effectiveness, service quality and efficiency while upholding transparency, propriety, public interest and the absence of conflicts of interest. Therefore, IWB is also very important to encourage state civil apparatus (SCA) to adapt rapid changes in the external environment and make maximum efforts to achieve organisational goals (Kim & Park 2017; Sabuhari et al. 2021), and SCA's IWB is able to increase competition and work professionally (Kamis et al. 2023).

To implement the principle of regional innovation, the role of regional apparatus organisation employees is needed to creatively address regional innovation problems that have not met the targets as stated in *Article 4 of the Government Regulation*, namely innovation in internal governance and implementation of regional government management functions, human resource management processes, public service innovation, public goods and services procurement processes and other regional innovations. This is a form of innovation that is the authority of the regional government, which aims to increase efficiency, effectiveness and responsiveness in meeting community needs.

This research is important to try to answer the problems faced by SCA in Maluku province by examining the influence of transformational leadership (TFL), work engagement (WE) on IWB and analysing the role of organisational justice (OJ) as a moderator. Several previous studies indicate that TFL plays a vital role in increasing IWB positively and significantly (Aditianto & Amir 2022; Aditya & Ardana 2016; Khudhair et al. 2022). This study uses OJ as a moderator to fill this research gap. Organisational justice is defined as an individual's perception of the extent to which decisions, processes and interactions within an organisation are perceived as fair, encompassing aspects of distributive, procedural, interpersonal and informational justice (Mustafa, Vinsent & Badri 2023; Wiseman & Stillwell 2022). The relationship between innovative behaviour and justice has yet to be adequately explored (Nazir et al. 2019), and there is a positive not significant direct effect in the influence of TFL on IWB and organisational innovation (Az Zahra & Etikariena 2024; Setiawan et al. 2021). However, several studies have found that OJ positively and significantly influences IWB (Kamis et al. 2023; Pakpahan et al. 2020). Khaola and Rambe (2021) underscore the critical role of fair treatment in the workplace and highlight the importance of leadership practices that promote justice and commitment to foster proactive and beneficial employee behaviours.

Employee WE in an organisation plays an important role in encouraging IWB, as individuals who are emotionally and cognitively engaged tend to be more motivated to generate creative ideas and new solutions. Research findings from Koroglu and Ozmen (2022); Salem et al. (2023); Syafitri and Etikariena (2023) confirm that high WE is positively correlated with increased IWB, which is essential for organisational competitiveness and adaptability in a dynamic business environment. Shuck, Adelson and Reio (2017) explained that employee WE refers to the intensity and direction of cognitive, emotional and behavioural energy. The influence of WE on IWB was studied using various models by previous studies without integrating with OJ variables; its influence and the underlying theory need to be updated. Organisational justice is integral to innovative behaviour, as without justice, engagement and innovation may falter (Jaboob et al. 2024; Lin et al. 2024). Given the discretionary nature of innovative behaviour, perceived unfairness in organisational processes may lead employees to withdraw psychologically and behaviourally, diminishing their willingness to innovate (Jašková 2017). Therefore, it is important for researchers to test theories collectively for empirical analysis and understanding, especially for SCA employees in North Maluku province. Although local government organisations have encouraged innovation through bureaucratic reform, studies on the correlation among TFL, WE, IWB and OJ in various provinces in Indonesia still need to be conducted. Despite growing interest in OJ and its role in promoting innovative behaviour, the moderating effect of OJ on the relationship between TFL, WE and IWB remains underexplored.

The research's findings provide two key theoretical contributions: firstly, extending and building theory with a model that confirms how OJ interacts with TFL and WE, which expands the IWB of employees in various in the Indonesian local government; and secondly extending social exchange theory by analysing the influence between characteristics of TFL, WE and IWB in a new global research model that explains OJ as a moderator.

## Literature review and hypothesis development
### Relationship between transformational leadership and innovative work behaviour

Innovative work behaviour involves generating, promoting and applying new ideas contributing to organisational performance, sustainability and long-term competitiveness. Recent studies affirm that IWB thrives when organisations implement high-performance work practices and foster supportive, inclusive environments that encourage individual creativity and collaboration (Gautam & Gautam 2024; Srirahayu, Ekowati & Sridadi 2023). Furthermore, strong empirical evidence shows that IWB significantly enhances organisational outcomes across sectors, including the public domain, by enabling adaptability and continuous improvement (Oh & Sabharwal 2024). In a holistic perspective, IWB includes not only individual creativity in generating ideas but also involvement in communicating, advocating and realising innovation in the work process (De Jong & Den Hartog 2010). Innovative work behaviour is influenced by various determinants, including TFL, organisational culture support and the level of job autonomy given to individuals (Amabile & Pratt 2016; Anderson, Potočnik & Zhou 2014). Therefore, to build a comprehensive IWB, a multidimensional approach is needed that combines cognitive, affective and conative aspects in encouraging individuals to become agents of change in their work environment.

Transformational leadership plays a pivotal role in fostering IWB within public sector organisations, primarily through its ability to inspire, intellectually stimulate and individually support employees. Leaders who embody transformational traits effectively create meaningful work environments that enhance employee engagement and creativity (Az Zahra & Etikariena 2024; Jun & Lee 2023). In the public sector context, especially regional government institutions, TFL is crucial for navigating bureaucratic constraints while fostering innovation (Nguon 2022; Wahyudi 2024).

Leadership research is not only a behavioural science but are present in various disciplines such as social, physics, management, psychology and others (Muller & Pelser 2022). Leadership has become an important part of the field of science that is discussed more widely and deeply (Bass & Bass 2008; Muller & Pelser 2022). Burns (1978) posited that TFL is a powerful tool for optimising an organisation's human resources to achieve predetermined goals and objectives. This leadership style not only encourages follower creativity through intellectual stimulation but also challenges assumptions and fosters an environment where leaders are willing to take risks and collect follower ideas, inspiring a culture of open-mindedness and innovation.

Transformational leadership inspires its followers to create a harmonious work environment to motivate them to increase creativity in the workplace. Transformational leaders are confident in themselves and believe in their potential to imagine and create a better organisational future. This TFL type consists of four factors: idealised influence traits, idealised influence behaviours, inspirational motivation and intellectual stimulation (Robbins & Judge 2015). Psychological factors have also been identified as important elements linking TFL to IWB. Research by Grošelj et al. (2021); Pradhan and Jena (2019) showed that employees who feel their work is meaningful, more competent and autonomous tend to be more innovative. This suggests that leaders who develop a goal-oriented work environment can enhance creativity. Azmi, Liana and Siregar (2023) asserted that when TFL is combined with a culture that values innovation, employee IWB increases significantly.

Leadership researchers propose that TFL can be used as a driver to facilitate IWB among employees (Sosik & Jung 2018). Transformational leadership plays a vital role in increasing IWB positively and significantly (Aditianto & Amir 2022; Azmi et al. 2023; Khudhair et al. 2022). Transformational leadership can be critical in increasing innovative employee work behaviour (Bibi et al. 2022). Thus, a successful leader can be a role model, be liked and create positive employee perceptions. Hence, we hypothesise the following:

**H1:** TFL has a significant effect on IWB.

### Relationship between work engagement and innovative work behaviour

Work engagement can be conceptualised through two dimensions: the physical and psychological vitality that employees bring to their work, expressed through enthusiasm, resilience and persistence when facing challenges. Identification, on the other hand, reflects a strong psychological connection to one's work, where individuals derive purpose and satisfaction from their roles. Recent studies reinforce WE, emphasising that engaged employees are more likely to demonstrate higher creativity, persistence and job performance (Bakker & Albrecht 2018; Mazzetti et al. 2023).

Furthermore, the relevance of engagement is increasingly linked to broader organisational outcomes. Gürbüz et al. (2024) highlight that WE is not only about increasing motivation but also about improving job design and leadership to strengthen energy and identification. Work enthusiasm, a mental state or behaviour that brings deep pleasure in employees to work diligently and consistently, is particularly noteworthy. It is this enthusiasm that drives individuals and groups to work towards achieving the institution's goals (Sastrohadiwiryo & Syuhada 2021). Dedication is characterised by a profound sense of commitment and involvement in one's work, where tasks are perceived as inspiring and evoke feelings of pride and enthusiasm, fostering a deep emotional connection to the job (Bakker & Schaufeli 2008). Complementing this, absorption describes a state of intense focus and positive immersion in work, where individuals become so engrossed in their tasks that time seems to pass rapidly, often making it challenging to disengage or stop working. Therefore, cultivating engaged employees is critical to sustainable organisational growth and innovation.

Work engagement has emerged as a significant predictor of IWB, as employees who are energetically and emotionally invested in their roles are more likely to generate, promote and implement new ideas. According to Sari, Yudiarso and Sinambela (2021), there is a moderate to strong correlation between WE and IWB, meaning that engaged employees tend to exhibit higher levels of creativity and problem solving at work. This finding is further supported by Yudiarso (2019), who found that civil servants' job engagement increases IWB, indicating its substantial role in driving innovation in the public sector. This evidence suggests that to enhance organisational innovation, especially in bureaucratic environments, management should prioritise strategies that stimulate employee engagement. In line with this, Alblooshi, Shamsuzzaman and Haridy (2021); Sethibe and Steyn (2015) found that there is a relationship between leadership and engagement, which in turn drives organisational innovation. In fact, WE has been found to significantly increase IWB among Indonesian regional civil servants (Kamis et al. 2023). This finding emphasises the importance of fostering employee engagement in organisations to promote a culture of innovation. Overall, the literature strongly supports the hypothesis that job engagement contributes significantly to IWB. Hence, we hypothesise the following:

**H2:** WE has a significant effect on IWB.

### Relationship between organisational justice and innovative work behaviour

Organisational justice refers to employees' perceptions of how they are treated fairly and equitably within the workplace, in alignment with established moral and ethical standards.

This sense of fairness includes distributive, procedural, interpersonal and informational justice (Mustafa et al. 2023; Wiseman & Stillwell 2022). This perception can significantly affect organisational performance, including employee commitment, job satisfaction and workplace harmony (Lyu 2016). By fostering a sense of justice, organisations can enhance trust, strengthen employee morale and promote positive behaviours, ultimately contributing to a more cohesive and productive work environment. Organisational justice emphasises the role of leadership decisions, perceptions of fairness and equality and the effects of justice on workplace dynamics, including the relationship of individuals with their work environment (Akram et al. 2016). It highlights how leadership decision-making and situational factors shape individual perceptions of justice within the organisation, ultimately influencing employee attitudes and behaviours. By examining these elements, OJ provides a framework for understanding how fairness is perceived and how it impacts organisational outcomes, such as trust, commitment and overall performance (Hussain et al. 2020). This perception of justice is closely related to increased trust in leadership, psychological safety and intrinsic motivation, all of which are important antecedents of IWB (Afsar & Badir 2016; Tamasevicius et al. 2025).

Empirical findings consistently support that OJ positively influences IWB (Afsar & Badir 2016; Pakpahan et al. 2020; Tamasevicius et al. 2025). For example, Afsar and Badir (2016) showed that there is a direct relationship between OJ and innovative behaviour, while Ghani et al. (2023) linked procedural justice to proactive behaviour, suggesting that fair procedures foster ownership and initiative, which are important traits for innovation. The theoretical foundation supporting these relationships draws on social exchange theory, which states that when employees perceive fair treatment, they reciprocate through positive discretionary behaviours, including innovation. This insight strengthens the hypothesis that OJ has a positive and significant effect on IWB, positioning justice as a strategic driver of innovation in human resource management. Hence, we hypothesise the following:

    **H3:** IWB is influenced by OJ.

    **H4:** OJ moderates TFL through IWB.

    **H5:** OJ moderates WE and IWB.

Based on the theoretical study and the proposed hypotheses, a conceptual framework for the research model was made as shown in Figure 1.

## Methods
### Sample and data collection

Empirically, we surveyed SCAs active in regional apparatus organisations in North Maluku province, one of the regions starting to develop in Indonesia. SCA was chosen as the unit of analysis of this study because it carries out work in the public service sector (government) with the vision, mission, goals and strategic objectives that the local government wants

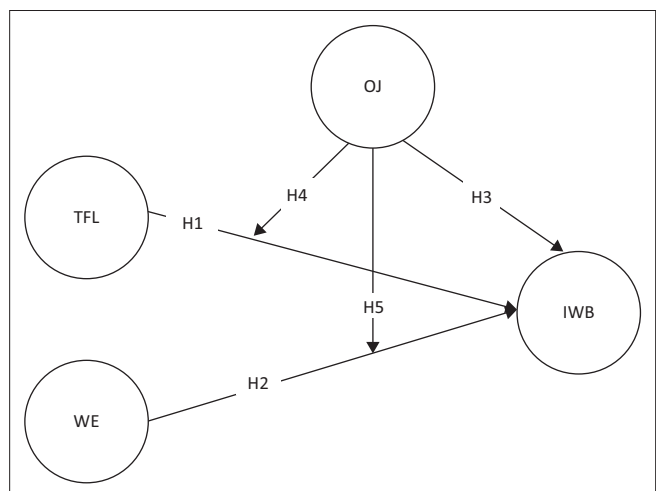

H, hypothesis; TFL, transformational leadership; OJ, organisational justice; WE, work engagement; IWB, innovative work behaviour.

**FIGURE 1:** Conceptual framework model.

to achieve. Therefore, transformational leadership, work involvement, IWB and OJ are critical to realising the vision, mission, goals and strategies the local government wants to achieve. Specifically, we use the innovation index assessment report 2022 in the category of less innovation (index value 33.11) and 2024 in the category of innovation (index value 58.47). Sampling in this study used non-probability with the purposive sampling method (Sekaran & Bougie 2013), to meet specific criteria (Wiyono 2011). The criterion in question employees have worked for more than 2 years. According to Hair et al. (2019), to determine the number of samples whose population is unknown with certainty, the indicators are multiplied by 5–10. The indicators total $64 \times 7 = 448$, which is the minimum number of samples for this study. Google Forms was used to send research questionnaires to SCA who were active in several less innovative and innovative regional apparatus organisations, as many as 500.

Out of 500 distributed questionnaires, 408 were successfully returned, yielding a substantial response rate of 81.6%. The demographic data revealed that 60.8% of respondents were male, while 39.2% were female. The majority of participants were aged between 31 and 50 years (64.8%), followed by those over 50 years old (23.5%) and a smaller proportion aged 21–30 years (11.7%). In terms of educational attainment, the majority of respondents held bachelor's or master's degrees (86.8%), while the remaining 13.2% comprised high school or vocational graduates and individuals with diplomas at the D3 or D4 level. These demographic insights provide a comprehensive understanding of the sample's composition and its representativeness.

### Measurement

The questionnaire made use of a five-point Likert scale (strongly disagree = 1 to strongly agree = 5). The questionnaire consisted of four sections: TFL, IWB, WE and OJ. The measurement framework in this study integrates and adapts multiple established indicators: TFL items were drawn from Bass and Avolio (1990), McCann, Morris and Hassard (2008) and Robbins and Judge (2015); IWB items from De Jong and Den Hartog (2010); WE items from Bakker and Schaufeli (2008) and Schaufeli et al. (2002) and OJ items from Lyu (2016) and Akram et al. (2016). These measurement items were subsequently refined and tailored by the researcher to align with the specific characteristics of SCA work in the local government sector.

### Data analysis and results

The data were analysed using statistical analysis of structural equation modelling-partial least squares (SEM-PLS), namely, tests of validity, reliability and hypothesis. The data analysis was conducted using the full version of SmartPLS 3.2.9 software, which was also employed to test the research hypotheses. The evaluation of the SEM-PLS involved assessing the measurement model by testing the discriminant validity, convergent validity and reliability of reflective indicators. Any indicators that did not meet the required loading factor thresholds were excluded from subsequent stages of analysis to ensure the robustness and accuracy of the results. Convergent validity value is used as a consideration of the loading factor > 0.70 and the average variance extracted (AVE) > 0.5 (Hair et al. 2017). Reliability in factor analysis is a key step to ensure the consistency of latent construct measurement by using indicators such as Cronbach's alpha, composite reliability and AVE, to ensure that the model has high stability and accuracy (Hair et al. 2017). Cronbach's alpha and composite reliability values are deemed reliable if both values exceed 0.7, as suggested by Hair et al. (2019). The analysis involved two stages of testing: the first focused on determining convergent validity, ensuring that the indicators within a construct were strongly correlated, and the second assessed discriminant validity to confirm that constructs were distinct and not excessively correlated with one another. Table 1 shows the first-stage analysis.

The results of the convergent validity analysis of the first stage of review from the outer loading score values obtained items that have a < 0.70, namely IWB 01 and 07 (0.471 and 0.679) and OJ 04 and 06 (0.614 and 0.598). These items can be stated as unable to measure the intended variables. Then, we continued the second-stage calculation by removing indicators with a value of less than 0.70. The second-stage calculation indicates that all indicators utilised have satisfied the validities, allowing the analysis to proceed with tests for convergent validity, discriminant validity and reliability using the embedded two-stage approach method (Hair et al. 2019), as illustrated in Figure 2.

The calculation results confirm that the dimensions employed to measure the latent variables meet the established criteria, adhering to the recommended guidelines (rule of thumb). The third-stage calculation indicates that TFL, IWB, WE and OJ meet the discriminant validity and reliability criteria. According to Hair et al. (2019), good reliability criteria are if the Cronbach's alpha value and composite reliability are more than 0.7 and for discriminant validity if AVE > 0.5. Detailed results are presented in Table 2 for further reference.

The inner model evaluates latent variables based on substantive theory and is tested using the goodness-of-fit (GoF) model to determine the contribution of independent variables to dependent variables. The $R^2$ (R-squared) value of the endogenous variable is part of the GoF analysis of the model, as shown in Figure 2. The $R^2 = 0.541$ indicates that TFL, OJ and WE collectively explain 54.1% of the variance in IWB, while the remaining 45.9% is influenced by factors outside the model. The subsequent model assessment incorporates moderation variables. In the realm of SEM-PLS, the presence of common method bias is suggested when the variance inflation factor (VIF) exceeds 3.3, which serves as a marker for pathological collinearity and raises the concern that the model may be adversely impacted by common method bias (Kock 2015; Sabuhari et al. 2023). Therefore, the VIF value must be lower than 3.3; then the model can be considered free from common method bias (Kock 2021). The results of the analysis showed that TFL to IWB = 2.388, WE to IWB = 2.643 and OJ to IWB = 3.206. The research model is free from common method bias. The analysis reveals that the path coefficient for the moderating effect of OJ on the relationship TFL with IWB is negative, as is the path coefficient of OJ on the relationship between WE and IWB.

Table 3 shows hypothesis testing results as to whether each hypothesis is accepted or rejected, which can be explained as follows.

The results of the data analysis explain each previously formulated hypothesis to determine acceptance or rejection. Table 3 shows that H1 posits that TFL influences IWB is accepted, indicating a significant effect. H2 suggests that WE has a positive and significant direct effect on IWB is providing strong empirical support. However, H3, which states that OJ affects IWB, is not supported. H4, which posits that OJ moderates the relationship between TFL and IWB, is also unsupported, suggesting that OJ does not serve as a moderator in this context. Finally, H5, which proposes that OJ significantly moderates the effect of WE on IWB, is rejected, providing no evidence of moderation. These findings indicate that further empirical exploration is needed to clarify the role of OJ in moderating these relationships. A visual representation of the influence between variables according to the tested hypotheses is provided in Figure 3.

## Discussion

This research aims to investigate the impact of TFL, WE and OJ on IWB among the SCA in North Maluku province, Indonesia. Additionally, it explores the moderating role of OJ in strengthening the relationship between TFL, WE and IWB, providing insights into how justice perceptions enhance innovation within the public sector. The analysis reveals that TFL and WE have a positive and significant direct impact on IWB. The effective implementation of TFL has been shown to significantly enhance SCA IWB, fostering innovative behaviour. The purpose of WE implementation is to boost productivity, retain employees within the organisation and

**TABLE 1:** Variables, dimensions and indicators (*Second order reflective*).

| Variable | Dimensions | CODE | Indicator | Loading factor | Cronbach's alpha dimension |
|---|---|---|---|---|---|
| TFL | Inspirational motivation | TFL 01 | Leaders motivate | 0.789 | 0.880 |
| | | TFL 02 | Leaders build self-confidence | 0.831 | |
| | | TFL 03 | Leaders provide confidence about the organisation's goals | 0.792 | |
| | | TFL 04 | Leaders generate enthusiasm | 0.860 | |
| | | TFL 05 | Leaders communicate about the work | 0.751 | |
| | Idealised influence | TFL 06 | Leaders are role models | 0.807 | 0.847 |
| | | TFL 07 | Leaders provide guidance | 0.766 | |
| | | TFL 08 | Leaders instil a sense of pride | 0.841 | |
| | | TFL 09 | Leaders are highly respected | 0.735 | |
| | Intellectual stimulation | TFL 10 | Leaders encourage creativity | 0.837 | 0.913 |
| | | TFL 11 | Leaders encourage innovation | 0.847 | |
| | | TFL 12 | Leaders are eager to listen to ideas | 0.771 | |
| | | TFL 13 | Leaders encourage rational or logical work problem-solving | 0.842 | |
| | | TFL 14 | Leaders solve problems from various perspectives | 0.802 | |
| | Individual consideration | TFL 15 | Leaders strive to improve employee self-development | 0.784 | 0.857 |
| | | TFL 16 | Leaders are fair to all employees individually and in groups | 0.742 | |
| | | TFL 17 | Leaders are willing to listen to employees' concerns | 0.785 | |
| | | TFL 18 | Leaders provide advice that is very important for employee development | 0.786 | |
| | | TFL 19 | Leaders value employees from different backgrounds | 0.832 | |
| IWB | | IWB 01 | Issues outside of routine work | 0.471 | - |
| | | IWB 02 | Efforts to improve work results | 0.725 | |
| | | IWB 03 | Searching for new methods and ways of working | 0.777 | |
| | | IWB 04 | Generating original solutions | 0.789 | |
| | | IWB 05 | New ways to complete work | 0.747 | |
| | | IWB 06 | New ideas get enthusiasm from colleagues (work team) | 0.820 | |
| | | IWB 07 | Efforts to convince others to support new ideas | 0.679 | |
| | | IWB 08 | Efforts to explain new innovation ideas in the work environment | 0.793 | |
| | | IWB 09 | Contributing to the implementation of new ideas | 0.720 | |
| | | IWB 10 | Developing new ideas | 0.706 | |
| WE | Actively participate in work | WE 01 | Taking part in various tasks | 0.871 | 0.771 |
| | | WE 02 | Active participation in carrying out tasks from the institution | 0.905 | |
| | | WE 03 | Self-involvement in current work | 0.797 | |
| | Affective commitment | WE 04 | Active personal involvement | 0.755 | 0.812 |
| | | WE 05 | This job is everything | 0.730 | |
| | | WE 06 | Regret negligence of work | 0.741 | |
| | Seeing work as something important for self-esteem | WE 07 | Able to carry out important tasks | 0.778 | 0.881 |
| | | WE 08 | Very strong bond with work | 0.798 | |
| | | WE 09 | This job is important for self-existence | 0.829 | |
| | Mental and emotional involvement | WE 10 | Developing ideas | 0.772 | 0.853 |
| | | WE 11 | Striving for the progress of the institution | 0.735 | |
| | | WE 12 | Interest is focused on work | 0.845 | |
| | | WE 13 | Work occupies the majority of one's daily time investment | 0.749 | |
| | Responsibility | WE 14 | Carrying out assigned tasks in a responsible manner | 0.817 | 0.782 |
| | | WE 15 | Putting tasks above everything else | 0.778 | |
| | | WE 16 | Completing main tasks before doing others | 0.795 | |
| OJ | Distributive justice | OJ 01 | Performance assessment of work done | 0.868 | 0.786 |
| | | OJ 02 | Appropriate grading | 0.862 | |
| | | OJ 03 | Assessing contributions to the organisation | 0.814 | |
| | | OJ 04 | Performance evaluations are justified | 0.614 | |
| | Procedural justice | OJ 05 | Appraisal procedures are based on accurate information | 0.728 | 0.826 |
| | | OJ 06 | There are other influences from appraisal procedures | 0.598 | |
| | | OJ 07 | Procedures are applied consistently | 0.757 | |
| | | OJ 08 | Procedures are free from response bias | 0.786 | |
| | | OJ 09 | Employees can appeal appraisals | 0.737 | |
| | | OJ 10 | Uphold ethical and moral standards | 0.784 | |
| | Interpersonal justice | OJ 11 | The superior treats subordinates politely | 0.815 | 0.880 |
| | | OJ 12 | The superior treats subordinates with consideration for self-esteem | 0.795 | |
| | | OJ 13 | The superior treats subordinates with respect | 0.868 | |
| | | OJ 14 | The superior refrains from making inappropriate comments | 0.708 | |
| | Informational justice | OJ 15 | The superior communicates honestly | 0.798 | 0.878 |
| | | OJ 16 | The superior explains procedures thoroughly | 0.859 | |
| | | OJ 17 | Supervisor's explanation of work procedures is logical | 0.872 | |
| | | OJ 18 | The superior communicates work in detail and in a timely manner | 0.883 | |
| | | OJ 19 | The superior is able to adjust his or her communication to the needs of the employees | 0.745 | |

*Source*: Research data tabulation (2024)
TFL, transformational leadership; OJ, organisational justice; WE, work engagement; IWB, innovative work behaviour.

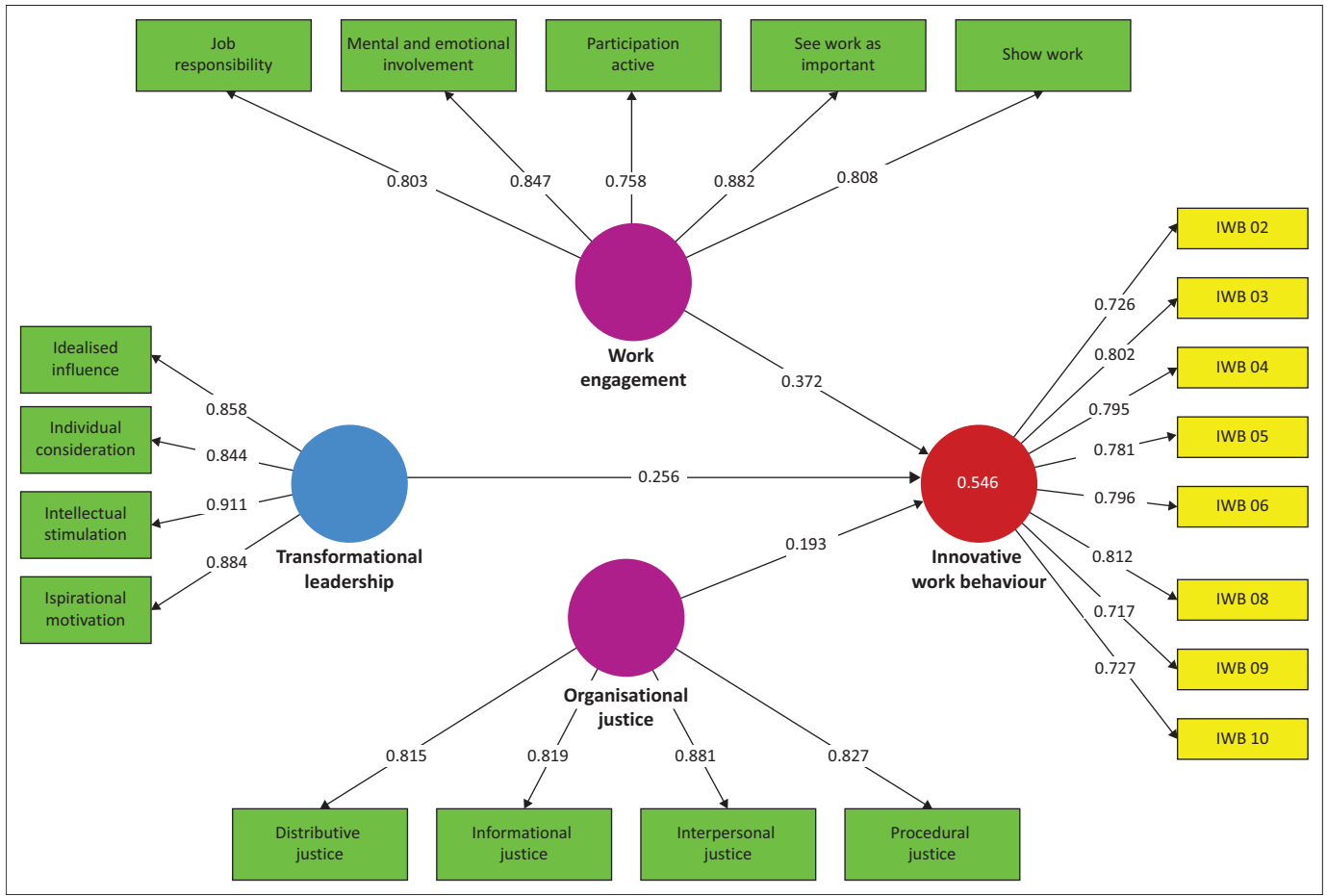

IWB, innovative work behaviour.
**FIGURE 2:** The third-stage of the embedded two-stage approach method.

**TABLE 2:** Reliability and discriminant validity.

| Variable | Cronbach's alpha | Composite reliability | AVE |
|---|---|---|---|
| IWB | 0.902 | 0.921 | 0.594 |
| OJ | 0.857 | 0.903 | 0.699 |
| TFL | 0.898 | 0.929 | 0.766 |
| WE | 0.879 | 0.911 | 0.674 |

*Source*: SmartPLS 3.2.9 calculation result output (2024)
AVE, average variance extracted; OJ, organisational justice; WE, work engagement; IWB, innovative work behaviour; TFL, transformational leadership.

improve community service delivery. By promoting WE, government organisations can increase productivity through innovative employee behaviour, thereby achieving both individual and organisational goals. This investigation elucidates that WE exerts a favourable and statistically significant influence on SCA's IWB, thereby supporting conclusions drawn from antecedent studies (Kim & Park 2017; Vithayaporn & Ashton 2019). These results suggest that when organisations actively engage employees directly in their work, then they can enhance their performance and adopt innovative behaviours.

Modern organisations strive to achieve ambitious long-term goals and showcase IWBs to demonstrate their competence and professionalism in a highly competitive environment (Sabuhari et al. 2021; Kim & Park 2017). This study primarily sought to evaluate how SCA considers the role of OJ in moderating the relationships between TFL, WE and IWB in Eastern Indonesia, using carefully selected indicators. Additionally, it aimed to assess how OJ influences TFL and its subsequent impact on IWB. The findings reveal that OJ does not significantly affect IWB, suggesting that the current implementation of OJ fails to enhance SCA's IWB. According to Gibson, Ivancevich and Donnelly (2009), despite OJ being a critical organisational factor, rooted in the theory of justice that employees typically evaluate their contributions to the rewards they obtain, however, in this research context that OJ does not significantly affect innovative behaviour. This indicates that, according to respondents' perceptions, OJ does not contribute meaningfully to employee innovativeness.

The researcher highlights OJ literature to explore its role as a source of motivation for employees' IWB. According to social exchange theory, fair treatment by organisations serves as a foundation for fostering positive employment relationships. However, this study's findings reveal that the variance in the observed aspects of OJ is minimal, indicating that OJ does not significantly motivate employees to engage in innovative behaviour. Specifically, SCAs in local government organisations operating in Eastern Indonesia perceive that OJ does not effectively drive their innovation. These results contrast with previous studies, which suggested that OJ positively contributes to enhancing employee IWB (Jnaneswar & Ranjit 2021; Kurniawan & Ulfah 2021; Sabuhari et al. 2023). This discrepancy

**TABLE 3:** Direct influence and moderating result.

| Hypothesis | Influence | Path coefficients | T statistics | p | Inner VIF values | Decision |
|---|---|---|---|---|---|---|
| H1 | TFL -> IWB | 0.230 | 3.385 | 0.001 | 2.388 | Accepted |
| H2 | WE -> IWB | 0.367 | 6.735 | 0.000 | 2.643 | Accepted |
| H3 | OJ -> IWB | 0.102 | 1.420 | 0.156 | 3.206 | Rejected |
| H4 | Moderating Effect1_*OJ*TFL -> IWB | −0.094 | 1.332 | 0.183 | 8.101 | Rejected |
| H5 | Moderating Effect2_*OJ*WE -> IWB | −0.015 | 0.301 | 0.764 | 7.962 | Rejected |

*Source*: SmartPLS 3.2.9 calculation result output (2024)
H, hypothesis; VIF, variance inflation factor; TFL, transformational leadership; OJ, organisational justice; WE, work engagement; IWB, innovative work behaviour.

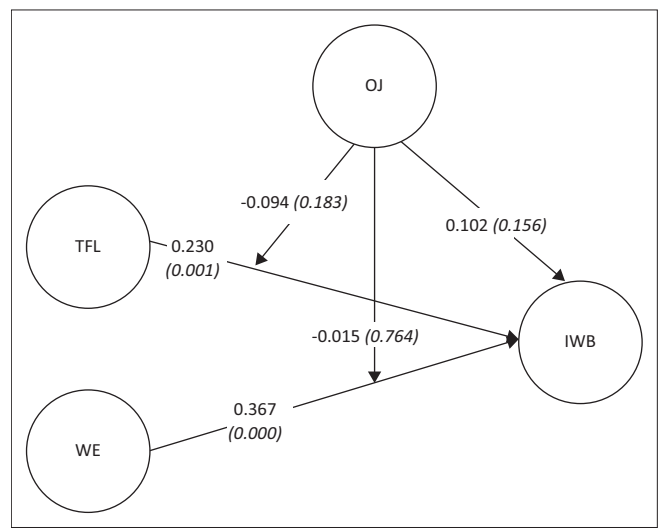

TFL, transformational leadership; OJ, organisational justice; WE, work engagement; IWB, innovative work behaviour.
**FIGURE 3:** Empirical confirmation that organisational justice does not play a moderating role.

underscores the need for further research to investigate contextual factors that may influence the relationship between OJ and IWB in different organisational and regional settings.

This research delineates the relations among TFL, WE and OJ concerning IWB while also examining the moderating influence of OJ. The empirical findings reveal a robust and statistically significant correlation between TFL and IWB, as well as between WE and IWB, thereby corroborating Hypotheses 1 and 2. Conversely, Hypotheses 3, 4 and 5 are not substantiated, as OJ does not exhibit a significant moderating effect on the relationship between TFL and IWB or WE and IWB. This shows that SCA in North Maluku Province can carry out its duties effectively to improve its innovative behaviour, even though it is not reinforced by organisational justice.

The study underscores the significance of WE and TFL as critical components to advance the innovative behaviour of employees in local governmental institutions. The results show that both WE and TFL have a positive and significant influence on the tendency of employees to demonstrate IWB, such as creating new ideas, daring to take risks and actively seeking alternative solutions in completing bureaucratic tasks. In the context of public organisations, the role of transformational leaders who can build an inspiring vision, provide motivation and empower employees is key in creating a psychological climate that is conducive to active employee engagement. When employees feel emotionally and intellectually involved in their work, the innovation potential naturally increases. Therefore, strengthening these two aspects is an important strategy in human resource management reform in the public sector.

On the other hand, the finding that OJ does not have a significant influence on IWB, either directly or as a moderator variable, provides interesting insights. Although justice remains a basic principle in organisational governance, the results of this study indicate that perceptions of justice are not necessarily the main driver of innovation in the context of local government institutions. This may be because of the dominance of psychological and motivational factors, such as inspiration from leaders and the level of personal involvement of employees in their work. Thus, innovation enhancement strategies should not only focus on structural aspects such as procedural and distributive justice but rather be directed at creating a work environment that encourages active participation, appreciation for new ideas and empowering leadership. These findings provide important contributions to the development of more contextual and evidence-based personnel policies in government bureaucracy.

The study strengthens the social exchange theory that emphasises the importance of reciprocal relationships between individuals and organisations and provides a theoretical basis for the development of human resource management based on trust, commitment and mutual benefit. In this context, TFL and WE reflect a form of positive exchange between the organisation and employees. When leaders show concern, provide inspiration and support individual development, employees feel appreciated and voluntarily respond with IWB as a form of reciprocity. Work engagement is also a manifestation of a trusting relationship that encourages employees to go beyond formal job demands and contribute more creatively to achieving organisational goals.

Furthermore, this theoretical implication underlies the importance of developing human resource management that does not only focus on administrative procedures but also on building relational and psychological qualities between superiors and subordinates. In the framework of modern human resource development, public sector organisations need to emphasise strategies that integrate social exchange values into Human Resource policies and practices, such as TFL training, innovative contribution-based reward systems and employee empowerment mechanisms. By ignoring OJ as a significant factor in driving innovation, this study suggests that human resource development must shift from a structural-formal approach to a relational-psychological approach. Therefore, social exchange theory not only explains the dynamics of employee innovative behaviour but also provides strategic direction for the design of Human Resource management systems that are capable of creating public organisations that are more adaptive, innovative and responsive to change.

# Conclusion

The results of this study confirm that TFL and WE are two key factors that significantly encourage the emergence of IWB in local government organisational environments. Leaders who can inspire, provide a clear vision and pay attention to individual development have proven effective in creating a work climate that supports innovation. Likewise, a high level of WE indicates that when employees feel emotionally, cognitively and physically involved in their work, they tend to be more proactive in generating new ideas. In contrast, the findings indicate that OJ, either directly or as a moderator variable, has no significant effect on IWB, indicating that perceptions of justice are not necessarily the main driver of innovation in the context studied. Therefore, managerial strategies and organisational policies should focus on strengthening the quality of TFL and increasing WE as a foundation for building an innovative culture in the workplace.

### Implications for local governments

The findings suggest that local governments should prioritise policies that enhance TFL capabilities and foster a work environment conducive to active employee engagement, as both TFL and WE significantly promote IWB. Efforts to strengthen TFL, particularly its visionary, inspirational and supportive dimensions, are essential to encourage innovation. Furthermore, policies should create opportunities for SCA participation and the appreciation of new ideas to bolster WE. Although OJ remains ethically essential, this study indicates that it does not significantly foster IWB in this context.

### Recommendations for future research

This study was conducted within a local government agency, where the work format is determined by the state, presenting conditions that differ significantly from those of private companies. As such, future research could evaluate the differences in IWB influenced by TFL, WE and OJ between government agencies and private sector organisations. To enhance the depth and scope of future studies, additional variables such as workplace happiness, knowledge sharing, and other relevant factors could be incorporated to provide a more comprehensive understanding of the dynamics influencing IWB in various organisational contexts.

## Acknowledgements

The authors would like to express their deepest gratitude to the Head of the Research and Community Service Institution, as well as the Rector of Universitas Khairun Ternate, for approving and funding this research through the 2023 Postgraduate Higher Education Superior Research Programme.

### Competing interests

The authors declare that they have no financial or personal relationships that may have inappropriately influenced them in writing this article.

### Authors' contributions

R. Sabuhari, executed and wrote up the study, while R. Soleman, was the study leader and provided supervision, M.M.S., J.F., and M.R., conceptualisation guidelines, methodology refinement, data analysis, editorial inputs and data tabulation, formal analysis, revision manuscript, and the writing of the English-language article.

### Ethical considerations

An application for full ethical approval was made to the Institute for Research and Community Services (LPPM) of Universitas Khairun sub-committee for research ethics in social sciences, and ethics consent was received on 05 February 2025. The ethics approval number is 56/UN44/L1/PG.03/2025.

### Funding information

The authors reported that they received funding from the Head of the Research and Community Service Institution (Contract Agreement Number: 038/PEN-PKUPT/PG.12/2023), as well as the Rector of Universitas Khairun Ternate, which may be affected by the research reported in the enclosed publication. The authors have disclosed those interests fully and have implemented an approved plan for managing any potential conflicts arising from their involvement. The terms of these funding arrangements have been reviewed and approved by the affiliated university in accordance with its policy on objectivity in research.

### Data availability

The data that support the findings of this study are available from the corresponding author, R. Sabuhari, upon reasonable request.

### Disclaimer

The views and opinions expressed in this article are those of the authors and are the product of professional research. It does not necessarily reflect the official policy or position of any affiliated institution, funder, agency or that of the publisher. The authors are responsible for this article's results, findings and content.

## References

Aditianto, P. & Amir, M.T., 2022, 'Pengaruh faktor kepemimpinan transformasional terhadap perilaku kerjainovatif melalui kesiapan untuk berubah sebagai mediator' [The influence of transformational leadership factors on innovative work behavior through readiness to change as a mediator], *Fair Value: Jurnal Ilmiah Akuntansi Dan Keuangan* 4(8), 3318–3326. https://doi.org/10.32670/fairvalue.v4i8.1381

Aditya, D.N.R. & Ardana, K., 2016, 'Pengaruh Iklim Organisasi, Kepemimpinan Transformasional, Self Efficacy terhadap Perilaku Kerja Inovatif' [The influence of organizational climate, transformational leadership, self efficacy on innovative work behavior], *E-Jurnal Manajemen Universitas Udayana* 5(3), 1801–1830.

Afsar, B. & Badir, Y., 2016, 'The mediating role of psychological empowerment on the relationship between person-organization fit and innovative work behaviour', *Journal of Chinese Human Resource Management* 7(1), 5–26. https://doi.org/10.1108/JCHRM-11-2015-0016

Akram, T., Jamal, M., Naqvi, H. & Feng, Y.X., 2016, 'The effects of organizational justice on the innovative work behavior of employees: An empirical study from China', *Journal of Creativity and Business Innovation* 2, 114–126, viewed 20 May 2025, from https://www.journalcbi.com/lander.

Alblooshi, M., Shamsuzzaman, M. & Haridy, S., 2021, 'The relationship between leadership styles and organisational innovation', *European Journal of Innovation Management* 24(2), 338–370. https://doi.org/10.1108/EJIM-11-2019-0339

Amabile, T.M. & Pratt, M.G., 2016, 'The dynamic componential model of creativity and innovation in organizations: Making progress, making meaning', *Research in Organizational Behavior* 36, 157–183. https://doi.org/10.1016/j.riob.2016.10.001

Anderson, N., Potočnik, K. & Zhou, J., 2014, 'Innovation and creativity in organizations', *Journal of Management* 40(5), 1297–1333. https://doi.org/10.1177/0149206314527128

Azmi, N., Liana, Y. & Siregar, Z.M.E., 2023, 'Transformational leadership, organizational culture and engagement on innovative behavior: A review', *International Journal of Business, Technology and Organizational Behavior (IJBTOB)* 3(3), 247–251. https://doi.org/10.52218/ijbtob.v3i3.280

Az Zahra, A.C. & Etikariena, A., 2024, 'The role of transformational leadership on innovative work behavior: A moderated-mediation study', *Jurnal Psikologi* 23(1), 81–96. https://doi.org/10.14710/jp.23.1.81-96

Bakker, A.B. & Albrecht, S., 2018, 'Work engagement: Current trends', *Career Development International* 23(1), 4–11. https://doi.org/10.1108/CDI-11-2017-0207

Bakker, A.B. & Schaufeli, W.B., 2008, 'Positive organizational behavior: Engaged employees in flourishing organizations', *Journal of Organizational Behavior* 29(2), 147–154. https://doi.org/10.1002/job.515

Bass, B.M. & Avolio, B.J., 1990, 'Developing Transformational Leadership: 1992 and Beyond', *Journal of European Industrial Training* 14(5). https://doi.org/10.1108/03090599010135122

Bass, B.M. & Bass, R., 2008, *The Bass handbook of leadership: Theory, research, and managerial applications*, 4th edn., Free Press, New York, NY.

Bibi, S., Khan, A., Hayat, H., Panniello, U., Alam, M. & Farid, T., 2022, 'Do hotel employees really care for corporate social responsibility (CSR): A happiness approach to employee innovativeness', *Current Issues in Tourism* 25(4), 541–558. https://doi.org/10.1080/13683500.2021.1889482

Burns, J.M., 1978, *Leadership*, Harper & Row, New York, NY.

De Jong, J. & Den Hartog, D., 2010, 'Measuring innovative work behaviour', *Creativity and Innovation Management* 19(1), 23–36. https://doi.org/10.1111/j.1467-8691.2010.00547.x

Gautam, P.K. & Gautam, D.K., 2024, 'High performance work practices for innovative work behavior: Mediating effect of workplace support and job embeddedness in IT-based service industry', *International Journal of Innovation Science*. https://doi.org/10.1108/IJIS-05-2023-0109

Ghani, B., Hyder, S.I., Yoo, S. & Han, H., 2023, 'Does employee engagement promote innovation? The facilitators of innovative workplace behavior via mediation and moderation', *Heliyon* 9(11), e21817. https://doi.org/10.1016/j.heliyon.2023.e21817

Gibson, J.L., Ivancevich, J.M. & Donnelly, J.H., 2009, *Organisasi dan manajemen: Perilaku, struktur, dan proses* [Organizations: Behavior, structure and processes], J. Wahid (ed.), Erlangga, Irwin McGraw-Hill, New York, NY.

Grošelj, M., Černe, M., Penger, S. & Grah, B., 2021, 'Authentic and transformational leadership and innovative work behaviour: The moderating role of psychological empowerment', *European Journal of Innovation Management* 24(3), 677–706. https://doi.org/10.1108/EJIM-10-2019-0294

Gürbüz, S., Schaufeli, W.B., Freese, C. & Brouwers, E.P.M., 2024, 'Fueling creativity: HR practices, work engagement, personality, and autonomy', *International Journal of Human Resource Management* 35(22), 3770–3799. https://doi.org/10.1080/09585192.2024.2429125

Hair, J.F.J., Hult, G.T.M., Ringle, C.M. & Sarstedt, M., 2017, *A primer on partial least squares structural equation modeling (PLS-SEM)*, 2nd edn., Sage Publications, Inc. Thousand Oaks, CA.

Hair, J.F., Risher, J.J., Sarstedt, M. & Ringle, C.M., 2019, 'When to use and how to report the results of PLS-SEM', *European Business Review* 31(1), 2–24. https://doi.org/10.1108/EBR-11-2018-0203

Hussain, I.A., Yunus, N.H., Ishak, N.A. & Daud, N., 2020, 'Effects of dimensions in organizational justice towards employee engagement', *International Conference on Management, Economics and Finance*, Hilton Hotel, Kuching, October 15–16, 2012, pp. 537–546, viewed 02 June 2025, from https://www.researchgate.net/publication/353260649.

Jaboob, A., Durrah, O., Mohd Ali, K.A. & Asha'ari, M., 2024, 'Organizational justice and innovative behavior at workplace: A comprehensive review', *Multidisciplinary Reviews* 7(9), 2024186. https://doi.org/10.31893/multirev.2024186

Jašková, I., 2017, *The relationship between organizational justice and innovative behaviour*, viewed 09 May 2025, from https://dspace.zcu.cz/items/e1bb9fee-9d76-41c1-8e7d-cf686ecc5dd0.

Jun, K. & Lee, J., 2023, 'Transformational leadership and followers' innovative behavior: Roles of commitment to change and organizational support for creativity', *Behavioral Sciences* 13(4), 2–19. https://doi.org/10.3390/bs13040320

Jnaneswar, K. & Ranjit, G., 2021, 'Organisational justice and innovative behaviour: Is knowledge sharing a mediator?', *Industrial and Commercial Training* 53(1), 77–91. https://doi.org/10.1108/ICT-04-2020-0044

Kamis, R.A., Al-Shami, S.A. & Sabuhari, R., 2023, 'The role of happiness at work in moderating the effect of work engagement and organizational justice on innovative work behaviour of Region SCA in Indonesia', *Journal of Human Resource Management – HR Advances and Developments* 26(2), 2–13. https://doi.org/10.46287/WVAP5643

Khaola, P. & Rambe, P., 2021, 'The effects of transformational leadership on organisational citizenship behaviour: the role of organisational justice and affective commitment', *Management Research Review* 44(3), 381–398. https://doi.org/10.1108/MRR-07-2019-0323

Khudhair, F.S., Rahman, R.A., Adnan, A.A.B.Z. & Khudhair, A.A., 2022, 'Study of the transformational leadership and organizational culture as predictors of employee creativity and innovation in the Iraq oil and gas service industry', *Zien Journal of Social Sciences and Humanities* 15, 34–50.

Kim, W. & Park, J., 2017, 'Examining structural relationships between work engagement, organizational procedural justice, knowledge sharing, and innovative work behavior for sustainable organizations', *Sustainability (Switzerland)* 9(2), 1–16. https://doi.org/10.3390/su9020205

Kock, N., 2015, 'Common method bias in PLS-SEM: A full collinearity assessment approach', *International Journal of E-Collaboration* 11(4), 1–10. https://doi.org/10.4018/ijec.2015100101

Kock, N., 2021, 'Harman's single factor test in PLS-SEM: Checking for common method bias', *Data Analysis Perspectives Journal* 2(2), 1–6, viewed 29 May 2025, from https://scriptwarp.com/dapj/2021_DAPJ_2_2/Kock_2021_DAPJ_2_2_HarmansCMBTest.pdf.

Koroglu, Ş. & Ozmen, O., 2022, 'The mediating effect of work engagement on innovative work behavior and the role of psychological well-being in the job demands–resources (JD-R) model', *Asia-Pacific Journal of Business Administration* 14(1), 124–144. https://doi.org/10.1108/APJBA-09-2020-0326

Kurniawan, D.T. & Ulfah, I.H., 2021, *The role of organizational justice in innovative work behavior of female employees in government institution*, Advances in Economics, Business and Management Research, pp. 9–18, Atlantis Press International B.V, viewed 24 May 2025, from https://www.atlantis-press.com/proceedings/bistic-21/125963886.

Lin, Q., Beh, L.-S. & Mohd Kamil, N.L., 2024, 'Perceived organizational justice and support facilitate employee innovation: A moderated mediation model of work engagement and empowerment', *Social Behavior and Personality: An International Journal* 52(3), 1–14. https://doi.org/10.2224/sbp.12961

Lyu, X., 2016, 'Effect of organizational justice on work engagement with psychological safety as a mediator: Evidence from China', *Social Behavior and Personality: An International Journal* 44(8), 1359–1370. https://doi.org/10.2224/sbp.2016.44.8.1359

McCann, L., Morris, J., & Hassard, J., 2008, 'Normalized intensity: The new labour process of middle management', *Journal of Management Studies* 45(2), 343–371. https://doi.org/10.1111/j.1467-6486.2007.00762.x

Mazzetti, G., Robledo, E., Vignoli, M., Topa, G., Guglielmi, D. & Schaufeli, W.B., 2023, 'Work engagement: A meta-analysis using the job demands-resources model', *Psychological Reports* 126(3), 1069–1107. https://doi.org/10.1177/00332941211051988

Muller, C.R. & Pelser, T.G., 2022, 'A proposed leadership skills development model for African FMCG business-networks: Super-Cube®', *South African Journal of Economic and Management Sciences* 25(1), a4303. https://doi.org/10.4102/sajems.v25i1.4303

Mustafa, M.J., Vinsent, C. & Badri, S.K.Z., 2023, 'Emotional intelligence, organizational justice and work outcomes', *Organization Management Journal* 20(1), 30–42. https://doi.org/10.1108/OMJ-08-2021-1322

Nazir, S., Shafi, A., Atif, M.M., Qun, W. & Abdullah, S.M., 2019, 'How organization justice and perceived organizational support facilitate employees' innovative behavior at work', *Employee Relations* 41(6), 1288–1311. https://doi.org/10.1108/ER-01-2017-0007

Nguon, V., 2022, 'Effect of transformational leadership on job satisfaction, innovative behavior, and work performance: A conceptual review', *International Journal of Business and Management* 17(12), 75. https://doi.org/10.5539/ijbm.v17n12p75

Oh, S. & Sabharwal, M., 2024, 'Fostering innovative work behavior to improve organizational performance', *Academy of Management Proceedings* 2024(1), a15748. https://doi.org/10.5465/AMPROC.2024.15748abstract

Pakpahan, M., Eliyana, A., Hamidah, Buchdadi, A.D. & Bayuwati, T.R., 2020, 'The role of organizational justice dimensions: Enhancing work engagement and employee performance', *Systematic Reviews in Pharmacy* 11(9), 323–332. https://doi.org/10.31838/srp.2020.9.49

Pooe, D. & Munyanyi, W., 2022, 'Delivering public value by selected government departments in South Africa – Perceptions of senior managers', *South African Journal of Economic and Management Sciences* 25(1), a3791. https://doi.org/10.4102/sajems.v25i1.3791

Pradhan, S. & Jena, L.K., 2019, 'Does meaningful work explains the relationship between transformational leadership and innovative work behaviour?', *Vikalpa: The Journal for Decision Makers* 44(1), 30–40. https://doi.org/10.1177/0256090919832434

Robbins, S.P. & Judge, T.A., 2015, *Organizational behavior*, Pearson Education, Inc., London.

Sabuhari, R., Jabid, A.W., Rajak, A. & Soleman, M.M., 2021, 'The role of organizational culture adaptation and job satisfaction in mediating effects of human resource flexibility on employee performance', *Jurnal Dinamika Manajemen* 12(1), 132–145, viewed 09 May 2025, from https://journal.unnes.ac.id/nju/jdm/article/view/27026/11991.

Sabuhari, R., Soleman, M.M., Adam, M.A. & Abdul Haji, S., 2023, 'Do adaptability and innovation speed matter in increasing sales of MSMEs during the COVID-19 pandemic?', *Journal of Economics, Business, & Accountancy Ventura* 26(1), 115–128. https://doi.org/10.14414/jebav.v26i1.2994

Salem, N.H., Ishaq, M.I., Yaqoob, S., Raza, A. & Zia, H., 2023, 'Employee engagement, innovative work behaviour, and employee wellbeing: Do workplace spirituality and individual spirituality matter?', *Business Ethics, the Environment & Responsibility* 32(2), 657–669. https://doi.org/10.1111/beer.12463

Sari, D.K., Yudiarso, A. & Sinambela, F.C., 2021, 'Work engagement and innovative work behavior: Meta-analysis study', in *Proceedings of the International Conference on Psychological Studies (ICPSYCHE 2020)*, Atlantis Press, Faculty of Psychology, Universitas Diponegoro, Semarang, October 20–21, 2020, pp. 359–366. https://doi.org/10.2991/assehr.k.210423.053

Sastrohadiwiryo, S. & Syuhada, A.H., 2021, *Manajemen tenaga kerja Indonesia [Indonesian Workforce Management]*, Bumi Aksara, Jakarta.

Schaufeli, W.B., Salanova, M., González-romá, V. & Bakker, A.B., 2002, 'The measurement of engagement and burnout: A two sample confirmatory factor analytic approach', *Journal of Happiness Studies* 3(1), 71–92. https://doi.org/10.1023/A:1015630930326

Sekaran, U. & Bougie, R., 2013, *Metode penelitian untuk bisnis: Pendekatan membangun keterampilan [Research methods for business: A skill-building approach]*, 6th edn., Wiley, New York, NY.

Sethibe, T. & Steyn, R., 2015, 'The relationship between leadership styles, innovation and organisational performance: A systematic review', *South African Journal of Economic and Management Sciences* 18(3), 325–337. https://doi.org/10.17159/2222-3436/2015/v18n3a3

Setiawan, B., Nugraha, D.P., Irawan, A., Nathan, R.J., & Zoltan, Z., 2021, 'User innovativeness and fintech adoption in Indonesia', *Journal of Open Innovation: Technology, Market, and Complexity* 7(3), 188. https://doi.org/10.3390/joitmc7030188

Shuck, B., Adelson, J.L. & Reio, T.G., 2017, 'The employee engagement scale: Initial evidence for construct validity and implications for theory and practice', *Human Resource Management* 56(6), 953–977. https://doi.org/10.1002/hrm.21811

Sosik, J.J. & Jung, D., 2018, *Full range leadership development*, Routledge, Oxfordshire.

Srirahayu, D.P., Ekowati, D. & Sridadi, A.R., 2023, 'Innovative work behavior in public organizations: A systematic literature review', *Heliyon* 9(2), 1–11. https://doi.org/10.1016/j.heliyon.2023.e13557

Syafitri, A. & Etikariena, A., 2023, 'The role of work engagement as moderator of perceived stress toward innovative work behavior', *Journal of Educational, Health and Community Psychology* 1(1), 78–106. https://doi.org/10.12928/jehcp.v1i1.25533

Tamasevicius, V., Pauliene, R., Casas, R., Russo, F., Puslyte, G. & Thrassou, A., 2025, 'The role of psychological empowerment in the impact of inclusive leadership on employee innovative behavior', *International Studies of Management & Organization* 55(2), 1–30. https://doi.org/10.1080/00208825.2025.2478351

Taylor, S.P., 2018, 'Innovation in the public sector: Dimensions, processes, barriers and developing a fostering framework', *International Journal of Research Science and Management* 5(1), 28–37.

Vithayaporn, S. & Ashton, S.A., 2019, 'Employee engagement and innovative work behavior: A case study of Thai Airways International', *ABAC ODI Journal Vision. Action. Outcome* 6(2), 45–62, viewed 09 May 2025, from https://assumptionjournal.au.edu/index.php/odijournal/article/view/3881/pdf.

Wahyudi, W., 2024, 'Transformational leadership and innovative work behavior: Mediating roles of organizational culture, self-efficacy, and work engagement in West Kalimantan State Polytechnics', *Jurnal Pendidikan Progresif* 14(2), 1162–1177. https://doi.org/10.23960/jpp.v14.i2.202484

Wiseman, J. & Stillwell, A., 2022, 'Organizational justice: Typology, antecedents and consequences', *Encyclopedia* 2(3), 1287–1295. https://doi.org/10.3390/encyclopedia2030086

Wiyono, G., 2011, *Merancang penelitian bisnis dengan alat analisis SPSS 17.0 & SmartPLS 2.0 [Designing Business Research with Analysis Tools SPSS 17.0 & SmartPLS 2.0]*, STIM YKPN, Yogyakarta.

Yudiarso, F., 2019, *Pengaruh work engagement terhadap Perilaku Kerja Inovatif pada Pegawai Negeri Sipil [The Influence of Work Engagement on Innovative Work Behavior in Civil Servants]*, Universitas Gadjah Mada, Indonesia.